\documentclass[a4paper,12pt,english,german]{article}
\usepackage{graphicx}

\begin{document}

{\bf 'Which Multiverse?': Some FAQ}

\bigskip

M. Dugi\' c\footnote{Department of Physics, Faculty of Science, Kragujevac, Serbia} and J. Jekni\' c-Dugi\' c\footnote{Department of Physics, Faculty of Science, Ni\v s, Serbia}

\bigskip

\noindent
Recently, we pointed out the possible inconsisntency in the very foundations of the Everett MWI (or a Multiverse) theory. Here, we place some emphasis on the very basic notions underlying our conclusion yet motivated by certain, recently raised clever observations in this regard.

\bigskip

{\bf 1. It is all about the standard quantum mechanics}

\bigskip

\noindent
Our analysis [1] is the standard and pure (nonrelativistic) quantum mechanics. On this basis, we start [1] with the general assumptions of MWI and we investigate some consequences.

We are interested in the standard linear canonical transformations (LCT) that, due to linearity, allow the inverse transformations likewise the direct quantization. The LCT we are interested in are {\it global}--they do not leave any degree of freedom of a World intact. As a consequence, a composite system $\mathcal{W}$ is differently decomposed into subsystems, e.g. $S+E = \mathcal{W} = S' + E'$. It {\it simultaneously}, {\it directly} and {\it without any additional requirements}  gives$^{N}$:

\smallskip

\noindent
a. the $\mathcal{W}$'s Hilbert state space $H$ decomposes as

\begin{equation}
H_S \otimes H_E = H = H_{S'} \otimes H_{E'},
\end{equation}

\noindent
b. the state $\vert \Phi \rangle$ of $\mathcal{W}$ in an instant of time $t_{\circ}$ can also be differently decomposed, in general, as:

\begin{equation}
\sum_i c_i \vert i \rangle_S \vert i\rangle_E = \vert \Phi \rangle = \sum_{\alpha} d_{\alpha} \vert \alpha \rangle_{S'} \vert \alpha \rangle_{E'}
\end{equation}

\noindent
c. the Hamiltonian $\hat H$ of $\mathcal{W}$ can also be decomposed as:

\begin{equation}
\hat H_S + \hat H_E + \hat H_{SE} = \hat H = \hat H_{S'} + \hat H_{E'} + \hat H_{S'E'}.
\end{equation}

To see the physical relevance of these notions remind the quantum theory of the hydrogen atom. There, the "electron ($e$)" and "proton ($p$)" are exchanged by the "center-of-mass ($CM$)" and the "relative particle ($R$)" systems (every possible textbook). As to the later, the absence of the interaction allows the variables separation and the exact solution to the Schrodinger equation for $R$. i.e. for the internal degrees of freedom of the atom.

So, our first claim reads: the standard and universally valid quantum mechanics allows, and sometimes (for the formal, mathematical reasons) requires the different decomposition of a composite system into subsystems, {\bf unconditionally}--as formally given by eqs. (1)-(3). ALL THE TIME WE DEAL WITH ONE AND ONLY ONE COMPOSITE SYSTEM, $\mathcal{W}$, for which we believe objectively to exist. E.g., $CM+R$ (but {\it not} "$e + p$") is what we observe for the hydrogen atom in the {\it bound} states, while:

\begin{equation}
\vert \psi_{qp}\rangle_{CM} \vert nlm_l m_s\rangle_R = H_{atom} = \sum c_i \vert i\rangle_e \vert i\rangle_p,
\end{equation}

\noindent
where the r.h.s. refers to the decomposition $e(lectron)+p(roton)$ that bears {\it entanglement} due to the non-zero (Coulomb) interaction; $\psi_{qp}$ may be a wave packet determined by the mean position and momentum values, $q$ and $p$, respectively.

{\bf In respond to Meeker} [2, 3]: the expression eq. (3) is unconditional (no "the time-like separate regions" is required in order eq. (3) to be valid), while referring to one and only one (yet a composite) physical system $\mathcal{W}$--such as the {\bf hydrogen atom}.

\bigskip

{\bf 2. In accordance with MWI}

\bigskip

\noindent
We adopt [1] the basic notion of MWI on decoherence as follows: {\it decoherence is a sufficient\footnote{Our best theories say it is also a necessary condition.} condition} for the (approximate) classical reality (and structure) of the World(s). The "pointer basis" is determined by the interaction [4], and there is not any conceptual or substantial distinction between the exact and the only-approximate pointer bases (the proof is simple yet to be presented elsewhere); e.g. the coherent state as present on the l.h.s. of eq. (4) is an element of an overcomplete, non-orthogonal basis states. Of course, the (realistic) approximate separability of the interaction [4] gives rise to approximate pointer basis.

Of course: it is possible that decoherence is not a sufficient condition--and this may "save" the MWI (cf. footnote 4).

{\bf Again, in respond to Meeker} [2, 3]: our analysis [1] equally refers to the exact as well as to the approximate (approximately orthogonal) pointer basis. [The later naturally takes into account the "nonideal" quantum measu\-re\-ments--in the modern literature recognized (and termed) yet as the "back action".]

\bigskip

{\bf 3. Some other questions}

\bigskip

i) We fix [1] the instant of time $t_{\circ}$ for several reasons. The answers below constitute the answers to the questions raised mainly by {\bf King} [5].

First, by fixing the instant of time, we uniquely define the quantum state of a unique $\mathcal{W}$--without the assumption of the unique time instant, eq. (2) loses its physical sense and meaning, when one can not even define the World's future dynamics. The analysis [1] refers equally to every possible World and to every possible instant of time. Second, the unique state of $\mathcal{W}$ determines the unique state for every decomposition of $\mathcal{W}$ in {\it every} instant of time. Third, the unique state of $\mathcal{W}$ in every instant of time determines its (future) dynamics--this is the issue of the {\it World's identity} within the consistent-history approach. On this basis, one can question [1] the splitting of the World $\mathcal{W}$ that starts in a properly defined instant of time $t_{\circ}$--the splittings referring to the different time instants are not interesting for us (except in the general context of the identity of the Worlds in the consistent-history approach to MWI that, cf. Section 2, we do not question).

\smallskip

ii) Looking for the "'maximally compatible basis' over all the interactions" is poorly stated yet intelligent a question [5].

MWI (and so do we [1]) deals with the "relative states". In this context, the 'maximally compatible basis' is just another set of the relative states--referring to {\it yet another decomposition} of the World. In other words: MWI does not exist without the Everett "relative states", while their introduction automatically calls for {\it another} (one out of the plenty of the possible) {\it decompositions of the World} into subsystems. So, there is not some "average" pointer basis, unless we abandon the very basis of MWI, i.e. the concept of the "relative states" and the corresponding (macroscopic) {\it structure} of the World.

\bigskip

{\bf 4. Justifying the main conclusion. Some open questions}

\bigskip

\noindent
It is easy to demonstrate the formal existence of the global LCTs implementing eqs. (1)-(3), i.e. to formally introduce the "Worlds" bearing the {\it similar global structure} (decomposition etc.) as the world we live in. So, the main argument stemming from eq. (2) [1] remains valid yet alternatively (more intuitively) expressed:

\noindent
{\bf either} {\it the decoherence effect is not sufficient (while we still assume it is a necessary condition) for the (approximate) classical reality of an open system}, {\bf or} {\it the MWI, while being sensitive to the variations of the Worlds'} global {\it  structure, should be abandoned}.

NOW, one may ask about the alternatives to MWI.

The only answer\footnote{That still assumes the decoherence effect is sufficient for the classical reality, while not {\it a priori} rejecting the "objective collapse".} we have (the details to be presented elsewhere) is as follows: following quantum mechanics, i.e. eqs. (1)-(3), the different decompositions (bearing yet the internal decoherence)
should be considered to be {\it equally realistic}--the relativity of the basic concept of 'system' [6, 7]. Interestingly enough, this NEW kind of the PARALLEL UNIVERSES (non-Everettian parallel universes (NEPU)), while having nothing in common with MWI, can\footnote{Assuming the presence of the intelligent beings in the alternative decompositions.} dynamically INFLUENCE each other\footnote{An action of an intelligent being in one decomposition, considered NOT as a spontaneous dynamics, inevitably lays "ghostly" a mark on the dynamics of all the other worlds.}! This is a consequence of the fact (cf. above) that we have the uniquely defined, non-splitting World in an instant of time $t_{\circ}$\footnote{It can not be overepmhasized: eq. (2) emphasizes the one and only one quantum state of $\mathcal{W}$ in every instant of time--i.e. the unique quantum state for every possible decomposition of $\mathcal{W}$ into subsystems.}.

\bigskip

$^N$The task of obtaining the r.h.s.  of eq. (2) from the l.h.s. and {\it vice versa} is in general a tough matter. For some simple models it can be shown that some LCT inevitably transform a separable into an entangled state [8] while the exact form of the entangled state is not easy to obtain; the reverse is an NP-problem. On the other side, in general, the LCT can change the symmetry of the composite-system's Hamiltonian. So, formally, investigating the transformations eqs. (1)-(3) is challenging an issue.

\bigskip

\noindent
{\bf NEPU Ontology.} There is only one Universe and only one Time as well as only one physical fundamental Law and the only one Universal quantum state of the Universe. [The Law may be the Schrodinger law, or the "objective collapse", or...] There are the different possible decompositions of the Universe into subsystems. Whilst mutually conditioned (by the "relative states" concept), the subsystems belonging to the same division should still be subject to decoherence. {\it Then} all such divisions (the NEPU) are {\it classically realistic} and {\it simultaneously existing}, i.e. sharing the same physical time.

\bigskip

[1] M. Dugi\' c and J. Jekni\' c-Dugi\' c, 2010, "Which Multiverse?",  arxiv:1004/1004.0148v1

[2] B. Meeker, 2010a, at

http://www.mail-archive.com/everything-list@googlegroups.com/msg18647.html

[3] B. Meeker, 2010b, at http://old.nabble.com/which-Multiverse--td28376271.html

[4] M. Dugi\' c, 1997, Physica Scripta {\bf 56}, 560

[5] S. P. King, 2010, http://old.nabble.com/which-Multiverse--td28376271.html

[6] M. Dugi\' c, J. Jekni\' c, 2006, Int. J. Theor. Phys. {\bf 45}, 2215

[7] M. Dugi\' c, J. Jekni\' c-Dugi\' c, 2008, Int. J. Theor. Physics {\bf 47}, 805

[8] De la Torre, A. C. et al, 2010,  {\it Europ. J. Phys.} {\bf 31}, 325

\end{document}